\documentclass[11pt,onecolumn,amssymb,nofootinbib,showpacs]{revtex4-1}
\usepackage{amsmath, amsthm, amscd, amssymb}
\usepackage{bm}
\usepackage{bbm}
\usepackage{graphicx}

\begin{document}

\title{Breaking quantum linearity: constraints from human perception and
cosmological implications}

\author{Angelo Bassi}
\email{bassi@ts.infn.it}
\affiliation{Dipartimento di Fisica Teorica,
Universit\`a di Trieste, Strada Costiera 11, 34014 Trieste, Italy.
 \\ Istituto Nazionale di Fisica Nucleare,
Sezione di Trieste, Strada Costiera 11, 34014 Trieste, Italy.} 

\author{Dirk-Andr\'e Deckert}
\affiliation{Mathematisches Institut der L.M.U. - Theresienstr. 39, 80333
M\"unchen, Germany}

\author{Luca Ferialdi}
\affiliation{Mathematisches Institut der L.M.U. - Theresienstr. 39, 80333
M\"unchen, Germany}


\begin{abstract}
Resolving the tension between quantum superpositions and the uniqueness of
the classical world is a major open problem. One possibility, which is extensively explored both theoretically and experimentally, is that quantum
linearity breaks above a given scale. Theoretically, this possibility is predicted by collapse
models. They provide quantitative information on where violations of the
superposition principle become manifest. Here we show that the lower bound on
the collapse parameter $\lambda$, coming from the analysis of the human visual
process, is $\sim 7\pm2$ orders of magnitude stronger than the original bound, in agreement with more recent analysis. This implies
that the collapse becomes effective with systems containing $\sim 10^4$-$10^5$
nucleons, and thus falls within the range of testability with present-day
technology. We also compare the spectrum of the collapsing field with those of known cosmological fields, showing that a typical cosmological random field can yield
an efficient wave function collapse.
\end{abstract} 
\maketitle

Schr\"odinger cat's states, namely quantum superpositions of states, each
containing a large number of particles, are part of the jargon of quantum
mechanics. Besides representing a fascinating challenge for experimental
physics~\cite{Nairz:01,Gerlich:07,Marshall:03}, due to their high sensitivity to external noises, they exemplify one
of the most controversial problems in modern physics~\cite{Bell:87,Seife:05}.
Quantum mechanics naturally allows for superpositions of macroscopic states to
occur, especially in measurement-like processes; however, such superpositions
are not part of our daily experience.

Several ways out have been proposed. Some re-interpret or reformulate the
quantum theory in a paradox-free way, others explicitly modify the quantum
dynamics. This second possibility is undertaken by models of spontaneous wave
function collapse~\cite{Ghirardi:86,Pearle:76,Bassi:03}.

According to these models, nonlinear and stochastic terms are added to the
Schr\"odinger equation. Their effect is to localize the wave function in space
in accordance with the Born rule. Physically, the collapse is driven by a
random field $w(t,{\bf x})$ which is supposed to fill space. A suitably tuned
coupling constant $\lambda$ makes sure that all quantum properties are
preserved at the microscopic scale~\cite{Adler:09}, while at the macroscopic
level classicality emerges.

As the noise $w(t,{\bf x})$ that causes the collapse is supposed to fill space,
its effects show up not only in laboratory experiments, but also at the cosmic
scale~\cite{Adler3:07}. This field interacts in fact with all matter in the
universe. At present there is no hint whether it can be identified with any
cosmological field considered in the literature, though some work has already
been done~\cite{Adler:08} and several suggested that it could have a
gravitational~\cite{Penrose:86,Karolyhazi:86,Diosi:89,Feynman:95} or
pre-quantum~\cite{Adler2:04} nature. Moreover, it is still not clear why it has
an anti-Hermitian coupling to matter, which is necessary to ensure the
collapse\footnote{With an Hermitian coupling, one would have a standard quantum Hamiltonian with a random potential; the equation would be linear and no suppression of quantum superpositions would occur.}.

Aim of this paper is twofold: the first goal is to provide a reasonable estimate for the collapse parameter $\lambda$. Such an estimate (its lower bound) is important, because it sets the scale at which a departure from the standard quantum behavior should be expected. This in turn is a crucial ingredient, in order to devise focussed experiments which would represent meaningful tests of quantum linearity. The second goal is to analyze the spectrum of the collapsing field as opposed to those of known cosmological fields. We aim at answering the question whether this field could reasonably have a cosmological origin, thus whether collapse models can be considered phenomenological models of an underlying theory, where the collapse emerges as the consequence of the interaction of quantum systems with a random field of nature.

There are several collapse models. The one that is commonly used in physical
applications is the Continuous Spontaneous Localization (CSL)
model~\cite{Ghirardi2:90}, which generalizes the original Ghirardi-Rimini-Weber
(GRW) model~\cite{Ghirardi:86} to quantum field theory. In the mass-proportional version of this model, the stochastically modified Schr\"odinger equation describing the dynamics is:
\begin{eqnarray} \label{eq:dfdsds}
| d \psi_t \rangle\!\!\! & = &\!\!\!  \left[ -\frac{i}{\hbar} H dt + \sqrt{\gamma} \int d^3x \left( M({\bf x}) - \langle M({\bf x}) \rangle_t \right) dW_t({\bf x}) \right. \nonumber \\
& & - \left. \frac{\gamma}{2} \int d^3x \left( M({\bf x}) - \langle M({\bf x}) \rangle_t \right)^2 dt \right] | \psi_t\rangle,
\end{eqnarray}
where $H$ is the standard quantum Hamiltonian of the system, while the other terms represent the stochastic and nonlinear modifications which induce the collapse of the wave function. The operator $M({\bf x})$ is an averaged mass-density operator:
\begin{equation}
M({\bf x}) = \frac{1}{m_N} \int d^3y\, g({\bf x} - {\bf y}) \sum_s m_s a^\dagger_s({\bf y}) a_s ({\bf y});
\end{equation}
in the above equation, the sum extends over particle species $s$ of mass $m_s$ with number density operator $a^\dagger_s({\bf y}) a_s ({\bf y})$, while $m_N$ is the mass of the nucleon and $g({\bf x})$ is a spatial correlation function, conventionally chosen equal to:
\begin{equation}
g({\bf x}) \; = \; \frac{1}{(2 \pi r_C^2)^{3/2}} e^{- {\bf x}^2/ 2r_C^2}.
\end{equation}
$\langle M({\bf x}) \rangle_t = \langle \psi_t | M({\bf x}) | \psi_t \rangle$ is the quantum average of the average mass-density operator, and $W_t({\bf x})$ is a family of standard independent Wiener processes, one for each point in space. Loosely speaking, the time derivative of the Wiener process: $dW_t({\bf x})/dt=w(t,{\bf x})$, can be interpreted as a Gaussian stochastic field filling space, whose time correlation function is a Dirac delta.

The fundamental (phenomenological) parameters of the model are the coupling constant $\gamma$ and the correlation length $r_C$, whose conventional values have been set equal to~\cite{Ghirardi2:90}: $\gamma \simeq 10^{-30} \text{cm}^3 \text{s}^{-1}$ and $r_C \simeq 10^{-5} \text{cm}$.  The collapse rate $\lambda$ is related to the above constant by the expression:
\begin{equation}
\lambda \; = \; \frac{\gamma}{8 \pi^{3/2} r_C^3} \; \simeq \; 2.2 \times 10^{-17} \, \text{s}^{-1}.
\end{equation}

The CSL model predicts that
superpositions of states separated by more than the correlation length $r_C$ of the noise are localized with a rate $\Gamma$ equal
to~\cite{Adler3:07}:
\begin{equation} \label{eq:rid}
\Gamma \; = \; \lambda\, n^2 N,
\end{equation}
where $n$ is the number of particles within a distance $r_C$, and $N$ is the
number of such clusters. This can be seen as follows. According to Eq.~\eqref{eq:dfdsds}, the decay of the
off-diagonal elements of the density matrix $\rho_t \equiv {\mathbb E}[|\psi_t\rangle\langle\psi_t|]$ (where ${\mathbb E}[\cdot]$ denotes the stochastic average) of a many-particle systems, which
gives a measure of the collapse process, is~\cite{Ghirardi2:90}:
\begin{equation}
\frac{\partial}{\partial t} \langle \bar{{\bf x}}' |\rho_t| \bar{{\bf x}}'' \rangle
= - \Gamma({\bar{{\bf x}}',\bar{{\bf x}}''}) \, \langle \bar{{\bf x}}' |\rho_t| \bar{{\bf x}}'' \rangle,
\end{equation}
where $\bar{{\bf x}}' \equiv {\bf x}'_1, {\bf x}'_2, \ldots {\bf x}'_N$ (and
similarly for $\bar{{\bf x}}''$). In the above equation, we have neglected the
standard quantum evolution. The decay function $\Gamma$ is defined as follows:
\begin{equation}\label{eq:blacsl}
\Gamma  = \frac{\gamma}{2}\sum_{i,j}\left[
G({\bf x}'_i-{\bf x}'_j) + G({\bf x}''_i-{\bf x}''_j) - 2G({\bf x}'_i-{\bf x}''_j)
\right],
\end{equation}
where the indices $i,j$ run over the $N$ particles of the system, and $ G({\bf
x}) = (4\pi r_C^2)^{-3/2}\exp[-{\bf x}^2/4r_C^2]. $

According to the equations above, when the particles in a superposition are
displaced by a distance $\ell = | {\bf x}' - {\bf x}''| \ll r_C$, their contribution to $\Gamma$ is negligibly small. Thus,
only superpositions with $\ell \geq r_C$ contribute to $\Gamma$ and trigger the
collapse of the wave function. In such a case, the formula tells that for
groups of particles separated (in each term of the superposition) by less than
$r_C$, the rate $\Gamma$ increases quadratically with the number of particles
while for groups of particles separated by more than $r_C$ it increases
linearly. Thus we have the following simplified formula for the collapse rate:
$\Gamma = \lambda \sum_i^N n_i^2$, where $n_i$ is the number of particles of the $i$-th group whose
mutual distance is less than $r_C$. Assuming for simplicity that $n_i$ is
constant for any $i$, we obtain Eq.~\eqref{eq:rid}. We note that the quadratic dependence of $\Gamma$ on the number of particles---which is absent in the original GRW model---is a direct effect of the identity of particles.

The conventional value of the collapse rate
$\lambda$ corresponds to a quantum-classical threshold of
$\sim 10^{13}$ nucleons~\cite{Ghirardi2:90}. Recently, a much stronger bound on
$\lambda$ has been proposed~\cite{Adler3:07}, namely $\lambda \simeq 2.2 \times
10^{-8 \pm 2}$ s$^{-1}$, corresponding to a threshold of  $\sim 10^5$ nucleons.
The underlying motivation is that the
collapse should be effective in all measurement processes, also those involving
only a small number of particles as it happens in the process of latent image
formation in photography, where only $\sim 10^5$ particles are displaced more
than $r_C$. In order for the collapse to be already effective at this scale,
one has to increase the conventional CSL value for $\lambda$ by $\sim 10^{9 \pm 2}$ orders of magnitude.

Both values are compatible with known experimental data~\cite{Adler:09}.
However, such a large discrepancy of $\sim 9$ orders of magnitude shows that there is no
general consensus on the strength of the collapse process and consequently on
the scale at which violations of quantum linearity can be expected to manifest.
This does not help the setting of future experiments, testing quantum
linearity. Here, we clarify the matter.

We consider the visual process, and in particular the fact that all human
perceptions are classical. Accordingly, any superposition reaching the eye must
be reduced before it is transformed into a perception in the
brain. This position has first been taken in~\cite{Aicardi:91}, in which authors assume that the reduction occurs at the very last step of the visual process, when the optical signal is transformed into a perception in the visual cortex. They show that by choosing $\lambda \simeq 2.2 \times 10^{-17} \text{s$^{-1}$}$ the collapse mechanism is strong enough to guarantee a classical perception. 
Anyway, it seems dubious that the collapse occurs in the
latest stages of the process. Otherwise, e.g., animals with a simpler visual
apparatus could perceive such superpositions which we consider rather unlikely.
We thus make the assumption that the eye acts as a detector, as also proposed
recently~\cite{Sekatski:09}. So we consider the case of a superposition of
few photons impinging and not impinging the retina, and we analyze for which
particles the relative displacement in the two branches of the superposition is
greater than $r_C$, thus contributing to the collapse. Since we want to set a lower bound on the collapse parameter, we take the worst case scenario, when only $\sim 6$ photons, corresponding to the threshold of vision~\cite{Hecht:41}, form the input signal. When such a small number of photons triggers the rods in the retina, the reaction time is
$\sim 100$ ms~\cite{Kandel:00}. 

Schematically, the whole process goes as
follows~\cite{Pugh:00} (see Fig. 1).
\begin{figure*}[b!]
\begin{center}
\includegraphics[width=0.9\textwidth]{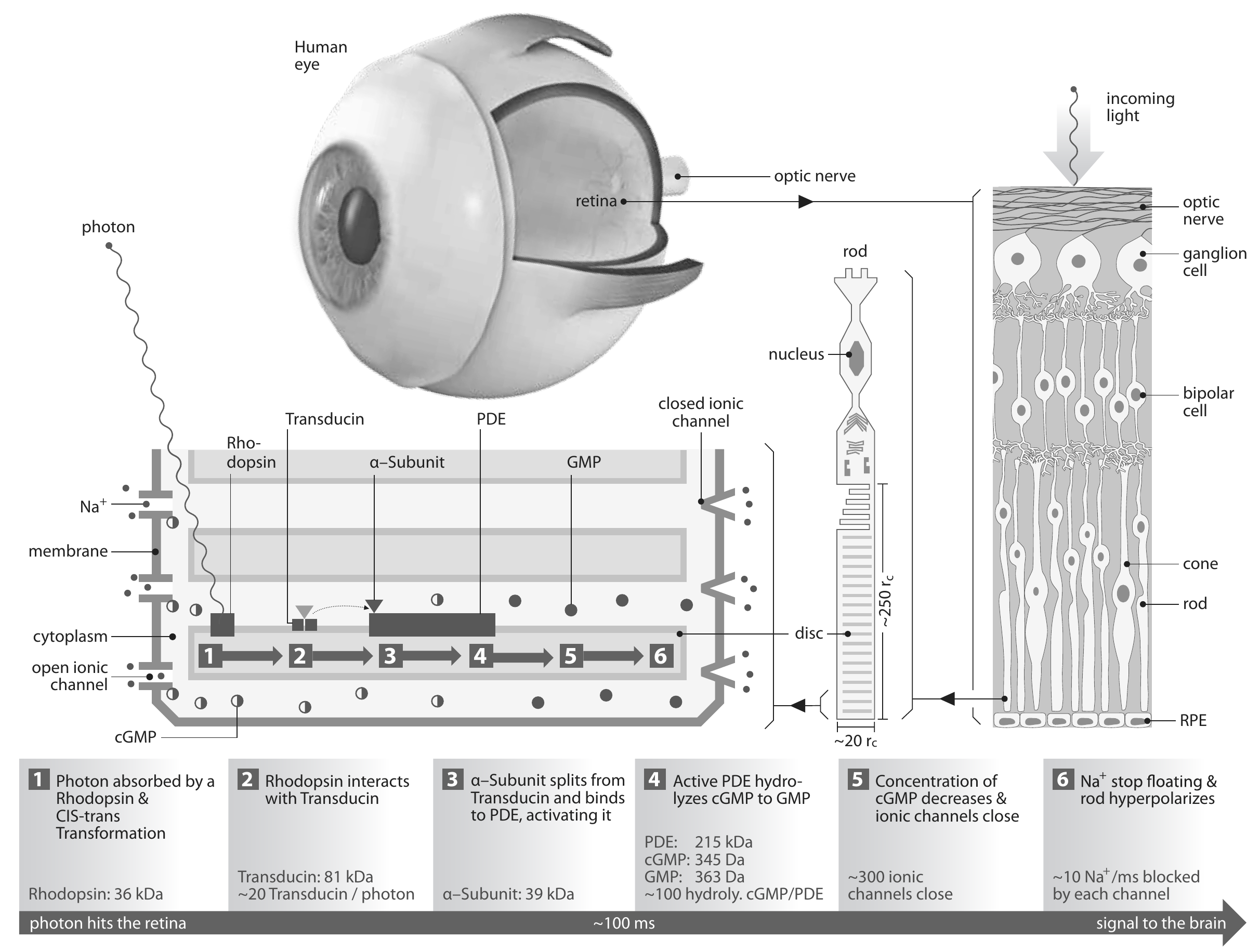}
\caption{Visual process in the human eye. The picture shows the most relevant
steps, focussing on how the incoming light is converted into an electric
signal.}\label{fig:1}
\end{center}
\end{figure*}
An incoming photon is absorbed by a rhodopsin in a rod disc, triggering a
cis-trans transformation. In this active form, the rhodopsin interacts with
$\sim 20$ transducins. The $\alpha$-subunit of each transducin splits from the
rest of the molecule, binding to a phosphodiesterase (PDE) and activating it.
During this chain of processes these molecules diffuse on the disc surface.
Next, each active PDE hydrolyzes $\sim 100$ molecules of cyclic guanosine
monophosphate (cGMP) to guanosine monophosphate (GMP). The decrease of
cytoplasmic cGMP concentration causes the closure of $\sim 300$ ionic channels
on the membrane of the rod~\cite{Nicholls:01}, each of which prevents $\sim 10$
Na$^+$ per ms to enter the rod~\cite{Rieke:98}. In this way the rod
hyperpolarizes and a graded potential is generated: this is the electric
signal, which is transmitted through the retina to the optic nerve. We now
estimate the contribution of each molecule to the reduction process.

Configurational changes of the molecules are small compared to $r_C$ and do not
contribute to the reduction process. Only relatively large displacements count.
The first relevant displacement is that of the $\alpha$-subunits of the
transducins, which remain attached to them or bind to the PDEs, depending on
whether a rhodopsin is active or not. Since the disc surface on which the $\alpha$-subunits can move has an area of $\sim 300\,r_C^2$, and reasonably assuming that the sub-units are uniformly distributed, then they are separated from each other by more than $r_C$.
These subunits then contribute with
$n_1^{\phantom{2}} = 3.9 \times 10^4$ (molecular weight of each subunit) and $N_1^{\phantom{2}} = 20$, providing a
factor of $n_1^2 N_1^{\phantom{2}} = 3.0 \times 10^{10}$ to Eq.~\eqref{eq:rid}.
The next relevant displacement is that of cGMP/GMP, which binds to the ionic
channels or diffuses in the cytoplasm, depending on its form. The concentration
of cGMP is $\sim 2 \mu$M~\cite{Pugh:00}, which amounts to $\sim 1$ molecule in
a volume of size $r_C$, so they are separated by more than $r_C$. There are $\sim 2000$ such molecules undergoing a
hydrolysation. This corresponds to $n_2^{\phantom{2}} = 363$ and
$N_2^{\phantom{2}} = 2000$, giving a contribution of $n_2^2 N_2^{\phantom{2}} =
2.6 \times 10^8$ to Eq.~\eqref{eq:rid}. 
This contribution is much smaller than the previous one\footnote{Being this contribution smaller than the previous one, even considering the case that cGMP is grouped in clusters of size $\lesssim r_C$, each containing $\sim 100$ molecules, would not affect the final result.}. 

The third relevant displacement is that
of the Na$^+$ ions. Since the surface density of ion channels is $500$ $\mu
\text{m}^{-2}$~\cite{Pugh:00}, there are about 5 channels within a distance
$r_C$. We remind that $\sim 300$ channels are involved in the process. In 100 ms $ \sim 10^3$ ions enter or fail to enter the membrane through
a channel~\cite{Rieke:98}. Since each channel brings the ions inside the molecules in groups of 3~\cite{Nicholls:01}, we distinguish two opposite situations: the more likely case case where the ions passing through a channel can be grouped in clusters of 3 molecules, distant less than $r_C$ (which means $n_3^{\phantom{2}} = 5\times3\times23$, $N_3^{\phantom{2}} =
60\times333$), and the extreme case where all ions passing through a channel are distant less than $r_C$ (thus $n_3^{\phantom{2}} = 5\times10^3\times23$, $N_3^{\phantom{2}} =
60$). Taking into account these two possibilities, the estimated contribution of the third step is  $n_3^2 N_3^{\phantom{2}}=2.4 \times 10^9 \div 7.9 \times 10^{11}$.
Summing up all these contributions, one obtains the overall factor $n^2 N$, which represents the effects of one photon impinging on the retina. Since the threshold of vision is $\sim$ 6 photons, we still have to multiply such a contribution by a factor of 6. 

When applying Eq.~\eqref{eq:rid} to estimate the value of the collapse parameter, one has to decide when a superposition can be said to have collapsed to a localized state. We use the criterion adopted in~\cite{Aicardi:91}, according to which a superposition has collapsed when $\Gamma \, t \simeq 10^2$, i.e. when one the two terms of the superposition is $e^{100}$ smaller than the other. Clearly this choice is arbitrary, though reasonable, and a value one order of magnitude smaller or bigger would still do the job. Taking this arbitrariness into account, and remembering that in our case
$t = 100$ ms, Eq.~\eqref{eq:rid} gives $\lambda\simeq5.0\times10^{-9\pm 1} \div 2.0\times10^{-11\pm 1}$
s$^{-1}$. A further contribution comes from the fact that the potential travels
through other cells in the retina, before entering the axon. Assuming that the
number of particles involved in each cell is of the same order as that for the
rod, and that an average of $\sim 3$ cells are involved, we obtain
$\lambda\simeq1.7\times10^{-9\pm1} \div 6.7\times10^{-12\pm 1}$ s$^{-1}$. We can summarize our result by saying that $\lambda \sim 10^{-10\pm2}$ s$^{-1}$.

This value is compatible with  $\lambda \sim 10^{-8\pm2}$ s$^{-1}$ first proposed in~\cite{Adler3:07}. Moreover, they both are quite surprisingly
close to the prediction coming from the analysis of the Schr\"odinger-Newton
equation in semiclassical gravity, according to which nonlinearities due to the
interaction with the gravitational field should appear already for systems with
$\sim 10^4$ nucleons~\cite{carlip}. At present it is not possible to further sharpen the numerical estimate of the lower bound. However, since $\lambda \sim 10^{-8 \pm 2}$
s$^{-1}$ makes sure that, either in measurement-like processes such as latent image formation, or in the visual process, the collapse of the wave function becomes effective, we consider this value in the rest of the manuscript. If this is true, then present day technology already allows for meaningful tests of quantum linearity,
since the collapse is predicted to be effective for systems containing $\sim
10^4$-$10^5$ nucleons being in a superposition\footnote{In systems like Bose-Einstein condensates, superconductors and superfluids, a much larger number of particles interact coherently. However, the typical center of mass displacement is smaller than $r_C$, thus collapse models do not predict a rapid loss of quantum coherence for these phenomena. Quantitative calculations show an agreement between the predictions of collapse models and experimental data~\cite{Adler:09}.} $\gtrsim r_C$.

A second related issue, already anticipated at the beginning of this letter, is whether the properties of the collapsing noise are compatible with general features of cosmological fields. If nonlinearities emerge when moving towards the macroscopic scale, and if these nonlinearities are associated to the interaction of physical systems with a random field filling space, then the most natural assumption is to suppose it to be a cosmological field. The first immediate question is whether such a field can reasonably have a cosmological nature, meaning with that, whether its most relevant properties (above all, its spectrum) are compatible with that of typical cosmological fields. The CSL model is not suitable
for this purpose as the spectrum of its noise is white, comprising all
frequencies with the same weight. Moreover, this noise represents an
unrealistic thermal field with infinite temperature.

Recently two generalizations of a simplified version of the CSL model (known as the QMUPL model~\cite{Diosi:89}) have been
thoroughly investigated, the first using a finite-temperature
noise~\cite{Bassi3:05}, the second using a colored noise with an exponential
correlation function~\cite{Bassi2:09}. These models are particularly useful for
deciding whether a cosmological fluctuating field with `typical' properties
can induce an efficient collapse. Equally important, they allow to follow
analytically the time evolution of wave functions. Here we investigate how the
collapse depends on the parameters of these models. We considered Gaussian
states
\begin{equation} \label{eq:gsol}
\phi_{t}(x) = \makebox{exp}\left[ - \alpha_{t} (x -
\overline{x}_{t})^2 + i \overline{k}_{t}x + \gamma_{t}\right]\,,
\end{equation}
which represent typical physical states. The spread in space of the wave
function, which measures the collapse, is:
$
\sigma_t = (2\sqrt{\alpha_{t}^{\text{\tiny R}}})^{-1/2},
$
the quantity $\alpha_{t}^{\text{\tiny R}}$ denoting the real part of
$\alpha_{t}$. The spread $\sigma$ has been plotted in Fig. 2 and 3.

Fig. 2 refers to the simplified CSL model.
\begin{figure}[t!]
\begin{center}
\includegraphics[width=0.9\textwidth]{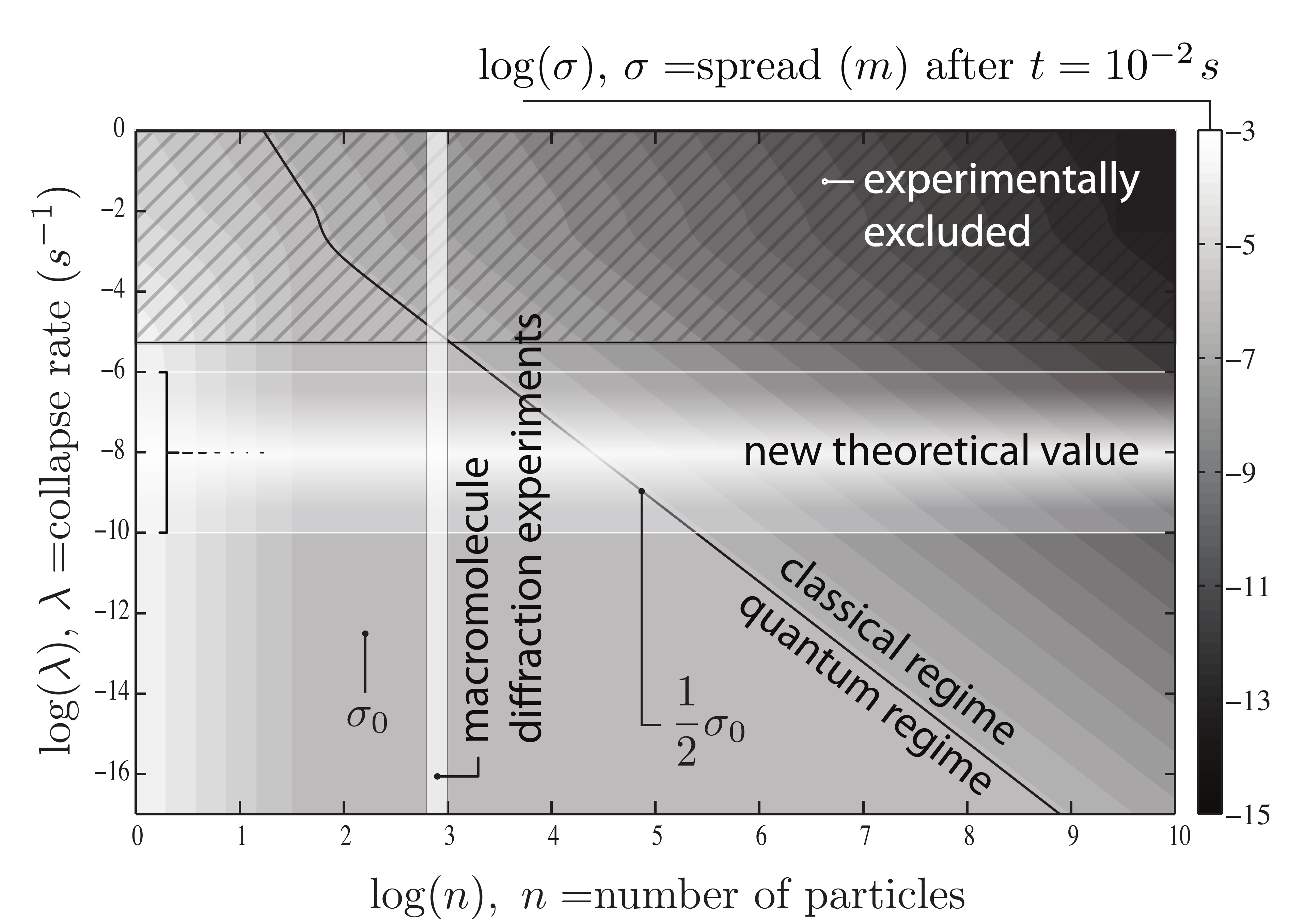}
\caption{Wave function spread $\sigma$ after a time $t = 10^{-2}$s, as a
function of particle number $n$ and collapse parameter $\lambda$, according to
the simplified CSL model. The initial value $\sigma_0 = 5 \times 10^{-7}$m
reproduces the geometry of the macromolecule diffraction experiments
of~\cite{Arndt:99,Nairz:01}. For few particles, and for small values of
$\lambda$, the spread increases just like in the standard quantum case.
Increasing $\lambda$, the collapse becomes stronger. The black line marks where
$\sigma$ has reduced to half its initial value, after the considered time. This can be taken as the
threshold from quantum (no collapse) to classical (collapse) regime. In the
case of the macromolecule diffraction experiments marked by the white strip,
values of $\lambda\lesssim10^{-5}$s$^{-1}$ imply that collapse models'
predictions agree with standard quantum predictions. This is consistent with
the analysis of~\cite{Adler3:07} for the CSL model. The proposed theoretical
bound, represented by the white region, shows that with an increase of the
particle number of $\sim2$-$3$ orders of magnitude, these experiments would represent a
crucial test for macroscopic quantum superpositions.}\label{fig:2}
\end{center}
\end{figure}
It shows how the spread of a wave
function with initial spread $\sigma = 5 \times 10^{-7}$ m changes after $t =
10^{-2}\text{s}$, as a function of the collapse strength $\lambda$ and of the
number of particles. The numerical values for $\sigma$ and $t$ mimic the setup
of the macromolecule diffraction experiments~\cite{Arndt:99,Nairz:01}, which up to now are the interferometric experiments involving the largest number of particles. As the picture shows, in order
for these experiments to become crucial tests of quantum linearity, the mass of
the molecules has to be increased of $\sim 2$-$3$ orders of magnitude. Recent
analysis~\cite{Gerlich:07} shows that such an improvement is already within
reach. Other crucial future experiments involve optomechanic
interferometers~\cite{Marshall:03}.

Fig. 3 (top row) shows the difference in the evolution of the spread as given by the
finite-temperature model and by the simplified CSL model.
\begin{figure}[t]
\begin{center}
\includegraphics[width=0.9\textwidth]{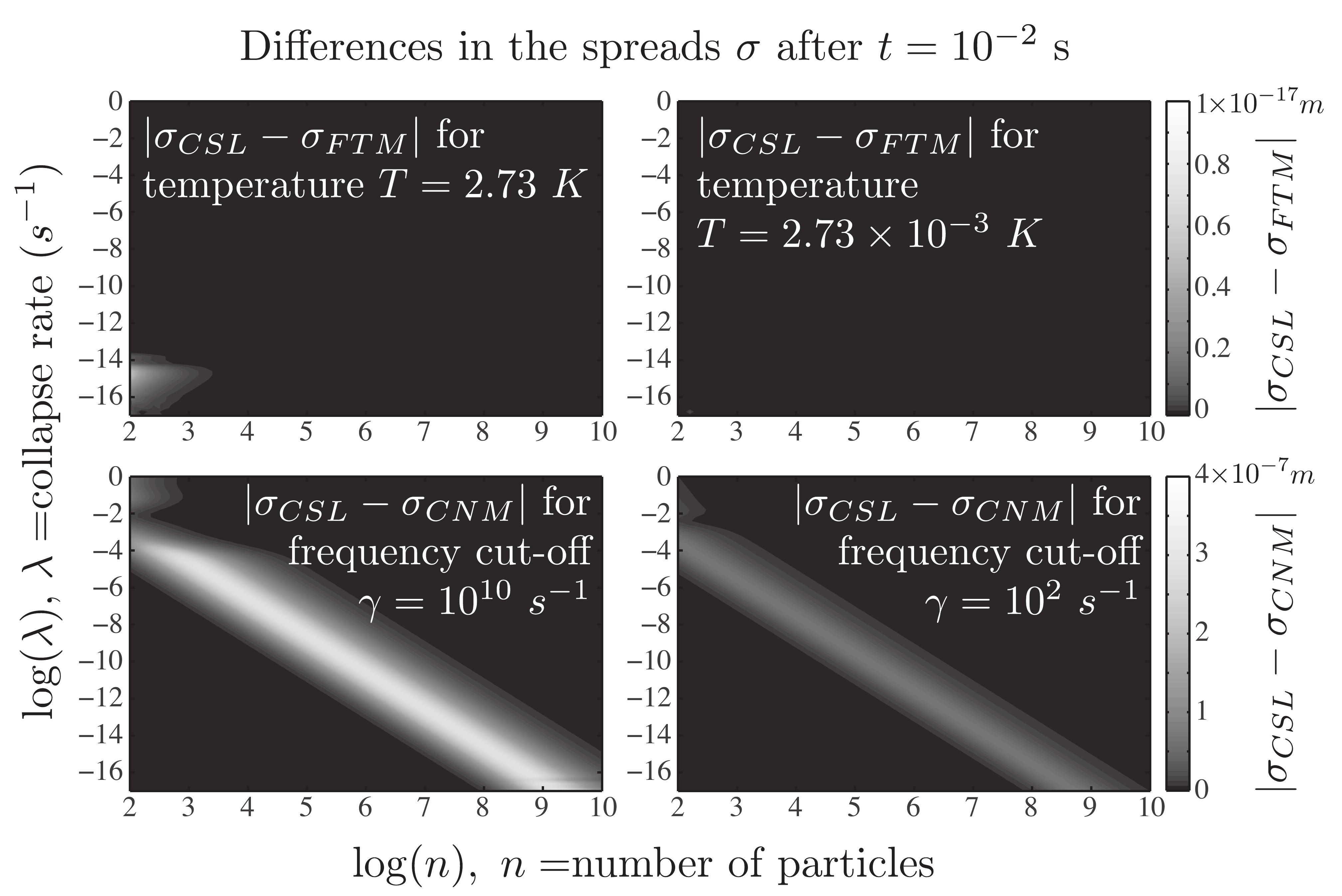}
\caption{Difference between the spread $\sigma$ predicted by the simplified CSL
model with that given by the finite-temperature model (FTM), for T = 2.73K (top row left) and T = 2.73 $\times 10^{-3}$K (top row right); and by the colored-noise model (CNM),
with cutoff at $10^{10}$Hz (bottom row left) and $10^2$Hz (bottom row right). As the color bars on the right show, the whiter the region, the greater the difference in the spreads. The initial spread $\sigma_0$ and the
elapsed time $t$ are the same as in Fig. 2. At lower temperatures or lower cutoffs the wave function tends to collapse more slowly, which results in a bigger
difference with respect to the simplified CSL model. Regarding the plots in the top row, the discrepancy
manifests in the lower left corner of the plot for $T=2.73\times 10^{-3}$K
which disappears for $T=2.73$K. Regarding the plots in the bottom row, it is manifest in the diagonal strip for
$\gamma=10^{2}$Hz which decreases for $\gamma=10^{10}$Hz. This diagonal feature
exactly lies on the ridge between the quantum and the classical regime as
displayed in Fig. 2. The large discrepancy there is due to the missing high
frequencies in the noise spectrum of the colored-noise model. Without these
high frequencies the colored-noise model is not able to reproduce the sharpness
of the ridge predicted by the simplified CSL model.}\label{fig:3}
\end{center}
\end{figure}
Two temperatures have been considered: $T = 2.73$K (top row right), as for the CMBR and $T =
2.73 \times 10^{-3}$K (top row left). In both cases the difference is negligible. This means
that even a rather cold thermal field, according to cosmological standards, can
induce an efficient collapse of the wave function as efficient as with the
standard CSL model. This does not mean that the two models are equivalent. On
the fast collapse scale ($1/\Gamma$) they agree while on the slower relaxation
scale, according to the finite-temperature model, the energy of any system
slowly decreases and thermalizes to that of the field, while according to the
CSL model it increases steadily.

Fig. 3 (bottom row) shows the difference in the evolution of the spread as given by the
simplified CSL model, and by the model with a noise having a frequency cutoff.
No high-frequency cutoff affects the collapsing properties of the model in an
appreciable way. These results can be compared with the behavior of typical
cosmological fields such as the CMBR, the relic neutrino background and the
relic gravitational background. The spectrum of the first two have a cutoff
(measured or expected) at $\sim 10^{11}$ Hz, while the spectrum of the third
one probably lies at $\sim 10^{10}$ Hz~\cite{Grishchuk:10}. All these cutoffs
as well as that of $\sim 10^{15}$ Hz proposed in~\cite{Bassi:03} for $w(t,{\bf
x})$ ensure a rapid collapse of the wave function. While the collapse is robust
over a large range of cutoffs, other effects, such as the emission of
radiation from charged particles, highly depend on the spectrum of the noise
correlator~\cite{Adler2:07}.

The message which can be drawn is that a cosmological field with `typical'
properties can induce an efficient collapse. A great challenge is to test the
existence of such a field. Our analysis, like that of~\cite{Adler3:07}, points towards a much stronger collapse
effect than previously thought. This means that crucial experiments testing
quantum linearity are within reach.

\vskip 0.3cm

\noindent \textsc{Acknowledgements.} We thank Dirk Aschoff for making Fig. 1,
and Dr. Paolo Codega for clarifying some technical issues related to the visual
process. L.F. acknowledges financial support from Della Riccia Foundations, and A.B. acknowledges partial support from MIUR, Italy, through the grant Prin2008.

\vskip 0.3cm

\noindent \textsc{Appendix 1: Finite temperature model.} According to this
model, the time evolution of $\alpha_t$ is given by~\cite{Bassi3:05}:
\begin{equation}
\alpha_t = -\frac{1}{2} \left[a + ib \tanh \left( \frac{\hbar}{n m_0}bt +
\kappa \right) \right],
\end{equation}
where $\kappa$ sets the initial condition $\alpha_0$, while
\begin{eqnarray}
a = - i \frac{\lambda \hbar n}{2 k_B T}\, \qquad b = \sqrt{|a|^2 + 2i
\frac{\lambda m_0}{\hbar} n^2}.
\end{eqnarray}
In the above expressions, $T$ is the temperature of the random field, $m_0$ is
the nucleon's mass and $k_B$ is Boltzmann constant. By taking the infinite
temperature limit (i.e. $a \rightarrow 0$), one recovers the time evolution as
given by simplified CSL model~\cite{Diosi:89}.

In using the formulas above we have replaced $n$ with $n^2$. This way we take
into  account that the collapse rate of the CSL model depends quadratically on
the number of particles whenever they are bounded in systems of size smaller
than $r_C$. This in particular is the case for macromolecules. \vspace{0.3cm}

\noindent \textsc{Appendix 2: Colored noise model.} Assuming an exponential
correlation function $D(t,s)$ for the noise~\cite{Bassi2:09}:
\begin{equation}\label{eq:expcorr}
D(t,s)=(\gamma/2)e^{-\gamma |t-s|}\,,
\end{equation}
where $\gamma$ gives the cutoff, the function $\alpha_t$ can be written as:
$
\alpha_t  = \mathcal{A}_t- (\mathcal{B}^2_t/4(\alpha_0+\mathcal{A}_t)),
$
where we have defined
\begin{eqnarray}
\mathcal{A}_t & = &-\frac{in m_0}{2\hbar\Theta_t}\sum_k\upsilon_k\left[\left(a_{\bar{k}}
\cosh\upsilon_{\bar{k}}t+b_{\bar{k}}\sinh \upsilon_{\bar{k}}t\right)
\cosh \upsilon_kt  \right.\nonumber\\
&+&\left.\left(d_k\sinh \upsilon_{\bar{k}}t-c \cosh \upsilon_{\bar{k}}t\right)
\sinh \upsilon_kt + d_{\bar{k}}\right]\,,\\
\mathcal{B}_t & = &-\frac{i n m_0}{\hbar\Theta_t}\sum_k\upsilon_k\left[a_{\bar{k}}
\cosh\upsilon_{\bar{k}}t+b_{\bar{k}}\sinh \upsilon_{\bar{k}}t \right.\nonumber\\
&+&\left. d_{\bar{k}}\cosh \upsilon_kt -c \sinh \upsilon_kt\right]\,,\\
\Theta_t & = & \sum_k\left[2c + \left(a_{\bar{k}} \cosh\upsilon_{\bar{k}}t+
b_{\bar{k}}\sinh \upsilon_{\bar{k}}t\right) \, \sinh \upsilon_k t\right.\nonumber\\
&+&\left. \left(d_k\sinh \upsilon_{\bar{k}}t-c \cosh \upsilon_{\bar{k}}t\right)
\cosh \upsilon_k t\right]
\end{eqnarray}
with $k\in\{+,-\}$ and $\bar{k}$ denotes the opposite sign. The new
coefficients are $a_{\pm} = \gamma \upsilon_{\pm}^3[\upsilon_{\pm}^2 \mp
\zeta]$, $b_{\pm} = \upsilon_{\pm}^2[\upsilon_{\pm}^4 \mp \gamma^2\zeta]$, $c =
\upsilon_{-}^3\upsilon_{+}^3$, $d_{\pm}=-\gamma
\upsilon_{\pm}^3\upsilon_{\mp}^2$, with:
\begin{equation}
\upsilon_{\pm} \; = \;\sqrt{\frac{\gamma^2 \mp \zeta}{2}}, \qquad\;\;
\zeta_{\phantom{\pm}} = \sqrt{\gamma^4-16i\gamma^2\frac{\hbar \lambda_{0}}{m_{0}}}.
\end{equation}
Also in this case, in using the above formulas we have replaced $n$ with $n^2$
to reproduce the fact that the collapse rate of the CSL model depends
quadratically on the number of particles.

\end{document}